\documentstyle[12pt]{article} 
 
\begin{document} 
 
\titlepage 
 
\def\expt#1{\mathord{<}#1\mathord{>}} 
\def\Yb{\bar Y} 
\def\Ub{\bar U} 
\def\Tb{\bar T} 
\def\Sb{\bar S} 
\def\Wb{\bar W} 
\def\e{\epsilon} 
\def\beq{\begin{equation}} 
\def\eeq{\end{equation}} 
\def\beqa{\begin{eqnarray}} 
\def\eeqa{\end{eqnarray}} 
\def\be{\begin{equation}} 
\def\ee{\end{equation}} 
\def\beq{\begin{equation}}   
\def\eeq{\end{equation}} 
\def\beqa{\begin{eqnarray}} 
\def\eeqa{\end{eqnarray}} 
\def\noin{\noindent} 
\def\grad{{\bf \nabla}} 
\def\pa{\partial} 
\def\kaps{{\kappa}^2} 
\def\Melev{ M^{11}} 
\def\bi{\begin{itemize}} 
\def\ei{\end{itemize}} 
\def\i{\item}

\begin{flushright} 
  QMW-PH-97-23 \\ 
\end{flushright}
 \vspace{1ex} 
 
\begin{center}\bf 
{\bf  GAUGINO CONDENSATION, MODULI POTENTIAL}\\ 
{\bf AND  SUPERSYMMETRY BREAKING}\\
 {\bf IN M-THEORY  MODELS }\\ 
\vspace{7ex} 
\rm
{\bf Zygmunt Lalak $^{\ast}$ and  Steven Thomas $^{\dagger}$ } 
\vskip 1.5cm
 $^\ast$  {\it Institute of Theoretical Physics \\ 
  University of Warsaw \\ 
   PL-00-681 Warsaw, Poland}\\ 
\vskip 1.5cm 
 $^\dagger$ {\it Department of Physics \\ 
         Queen Mary and Westfield College \\ 
        Mile End Road, London E1 4NS, UK } \\
\vskip 1.5cm 

\vspace{6ex} 

{\bf Abstract}

\end{center}

 \vskip 0.5cm
 We  derive the explicit form, and discuss some properties of the  
 moduli dependent effective potential
arising from M-theory
compactified on $M_4 \times X\times S^1 / Z_2 $,  when one of the boundaries 
supports a strongly interacting gauge sector and induces 
gaugino condensation.
We discuss the relation between the explicit gaugino condensate and 
effective superpotential formulations and find  interesting 
differences with respect to the situation known from the weakly 
coupled heterotic string case.
The moduli dependence of the effective potential turns out to be 
more complicated than expected, and perhaps offers new clues to 
the stabilization problem.

\newpage 
 
\section{Introduction} 
 
Lack of understanding of the mechanism of realistic supersymmetry breaking  
is the crucial missing ingredient  in 
supersymmetric theories of fundamental forces and the obstinate roadblock  
 in the supersymmetric unification programme.  
The other troublesome problem in these schemes is  
the apparent lack of unification between gauge and  
gravitational couplings. Although the low energy 
 considerations suggest unification  
of gauge couplings in supersymmetric extensions of the Standard Model at the  
mass scale $M_{GUT} < M_{PLANCK}$, the Newton constant scaled with energy  
as $E^2$ comes out to be too small to unify with the other  
couplings at $M_{GUT}$ 
in standard  scenarios, cf. \cite{polch}. The way to avoid this  
trouble has been suggested  
by Witten, \cite{strw}, in the framework of the field theoretical limit of the strongly coupled 
string theory, usually referred to as field theoretical limit of the M-theory.  
Witten and Horava \cite{whor} have argued that the effective low energy field theory  
 stemming from M-theory and describing the low-energy limit of the  
strongly coupled 
heterotic string is the 11-dimensional supergravity on the  
manifold with boundary, 
which couples to 10d supersymmetric  
Yang-Mills theories living on the components of 
the 10-dimensional boundary. 
 
 In the model of Witten and Horava  
there are two components  
of the boundary each containing one $E_8$ super-YM sector. The two sectors 
communicate  
with each other through the gravitational mediation,  and, interestingly 
enough, 
 in the regime where the requirement of the extended unification in the 
visible world is 
 fulfilled, one of them is more strongly coupled than the other. This 
suggests that we are 
 given in a natural way  
the  much desired 4d supergravity hidden sector with the strongly coupled $E_8$  
group, where we expect gauge fermions, gauginos, to condense at the dynamically 
generated scale $\Lambda_8$. Given this observation, one can hope that this condensation  
can be the natural source of supersymmetry breaking in the model (in analogy with  
classic considerations in the weakly coupled heterotic string models,
\cite{revs}, \cite{dine}).  
Indeed, the work of Horava, \cite{horava}, shows, that hidden sector gaugino condensation  
breaks 11d supersymmetries, and does it in a very interesting and nontrivial way.  
However, this phenomenon deserves further investigation.  
 
For instance, since the visible  
sector is separated from the condensing sector, it is not obvious what is the specific  
form of the local operators violating supersymmetry in  
the visible sector. The hypothesis of Antoniadis and Quiros \cite{anton} 
(cf. also Dudas and Grojean \cite{dud})
 puts forward the  
version of the Scherk-Schwarz mechanism as the effective description of the gaugino condensation.  
The Scherk-Schwarz mechanism however implies a very specific contribution to the potential 
energy density. One should be able to understand this contribution  
starting from the  
fundamental Lagrangian with multifermionic  
terms, which hasn't been attempted yet.  
Next, the picture of Horava assumes the stiff condensate, which is not a dynamical variable  
subject to backreaction of other dynamical degrees of freedom. In the reality this is not correct, as the condensate is going to be a function of several moduli fields, actually forms  
a dynamically generated potential for these fields, and its actual magnitude should be  
determined upon the minimization of the potential over the moduli space.  
This brings back  
the problem of moduli stabilization and its possible connection  
to supersymmetry breaking mechanism - it would be very disappointing if there wouldn't be any.  
We stress here the fact that the condensate is in fact a dynamical variable, following the  
classic work of Dine et. al \cite{dine}, as it has important consequences 
in the weakly coupled heterotic string. 
 
 There it turns out that, if one forgets about T-duality, the dynamical condensate  
tends to adjust itself in such a way that supersymmetry remains unbroken,
and moduli run away into ultra-weakly coupled regime. In those models one 
needs a  specific T-dual  
superpotential for the moduli, and usually some subsidiary tools like chiral matter in the 
 hidden sector,  to get  susy breaking in the direction of one of the moduli.  
Here, as the breaking is associated with boundary conditions which project away solutions to the Killing spinor equations, one would expect that once the condensate  
is nonzero, supersymmetry is broken, but still there remains the question
about the magnitude of
 the condensate (and the gravitino mass which is related to it) and about the 
form of the effective potential for the moduli.

Finally, there is the associated problem of the general form of the effective Lagrangian  
seen at low energies in the 4-dimensional observable sector. This sector contains at least  
two pieces 
\beq 
{\cal L} = {\cal L}_{obs} (Q, M) + {\cal L}_{moduli}(M)   
\eeq 
where $Q$ denotes observable fields and $M$ are moduli - which don't have to be  
much heavier than 1 TeV. It is interesting to note, that usually authors tend to think even  
in the context of M-theory models that this Lagrangian should  
be derivable from some effective 4d supergravity, 
which then - from the 4d point of view - breaks down spontaneously,  
and that there must exist  
a 4d superpotential for moduli M. However, in the present context, when
 supersymmetry breaking in 11d 
(or 5d after obvious compactification) arises from boundary conditions in
11th (or 5th) dimension, what one can naturally expect is explicit
supersymmetry in 5d, so for instance the existence of the effective
superpotential for moduli is not really obvious. This seems to be an
important point, as one tends to describe the supersymmetry breaking at low energies  
in terms of F-terms, and the nonvanishing ones are expected to arise exactly in the moduli  
sector. To discuss seriously and reliably the F-terms  
one needs the form of the effective moduli potential (and superpotential). 
This is the fundamental question  
 which we raise and attempt  
to discuss in this paper.  
 
To perform our task we shall used mixed techniques, reducing the fermionic terms from  
10 dimensions and then deducing moduli dependence of the condensate
 through the  
gauge coupling dependence of various scales, but also trying to construct directly the  
effective superpotential for moduli.

Our paper is structured as follows: in section 2 we consider
compactification of the  M-theory effective field theory, 
corresponding to strongly coupled $E_8 \times E_8 $ heterotic strings, 
in which gaugino condensation occurs on one of the two boundaries, 
in the presence of non vanishing $G_{11ABC} $. The emphasis is on 
the computation of the moduli dependence of the resulting 
four dimensional effective action. In section 3 we consider to what extent
such an effective potential can be reconstructed from standard $N=1, d= 4$
supergravity with a moduli dependent superpotential $W(S,T) $ whose form 
has already been proposed in the recent literature. We shall achieve  only
partial success in this task, encountering a number of problems which 
 appear, at least, to have a common origin. Some qualitative remarks 
concerning the possible stabilization of moduli expectation values are
 also made.  
With the various difficulties referred to above in mind,  we make some preliminary comments 
in section 4 concerning the origin of soft supersymmetry breaking masses
in potentially realistic models. We end with conclusions.

\section{Gaugino condensation and effective potential}
 
To start with let us recall the form of the M-theory Lagrangian 
constructed by Horava and Witten \cite{whor}, which is given by $L_S + L_B
$ where 
\beqa\label{eq:1} 
 {L_S} & = & \frac{1}{\kaps}\int_{\Melev} d^{11}x \, \sqrt{g} \, \{ 
 -\frac{1}{2} R - \frac{1}{2} {\bar{\Psi}}_I \Gamma^{IJK} D_J  {\Psi}_K 
 - \frac{1}{48} G_{IJKL} G^{IJKL} \cr 
&&\cr 
& - & \frac{\sqrt{2}}{384} ( 
 {\bar{\Psi}}_I \Gamma^{IJKLMN} \Psi_N + 12 {\bar{\Psi}}^J \Gamma^{KL} \Psi^M ) 
\,( G_{JKLM} + {\hat{G}}_{JKLM} ) \cr  
&&\cr 
 &-& \frac{\sqrt{2}}{3456} \epsilon^{I_1...I_{11}} C_{{I_1}{I_2}{I_3}} 
  G_{I_4 ...I_7} G_{I_8 ...I _{11}} ) 
\eeqa

\beqa\label{eq:2} 
 {L_B} & = &  \frac{1}{2 \pi {(4 \pi \kaps )}^{2/3} }   \sum_{m=1}^2 
 \int_{M^{10}_m} d^{10} x  \, \sqrt{g} \, ( -\frac{1}{4} {\rm Tr}\, F^{m}_{AB} 
F^{mAB} - \frac{1}{2} {\rm Tr}\, \bar{\chi}^{m} \Gamma^A D_A ({\hat{\Omega}
} )\, 
\chi^{m} \cr 
&&\cr 
 &- & \frac{1}{8} {\rm Tr}\, {\bar{\Psi}}_A \Gamma^{BC} \Gamma^A 
 (F^{m}_{BC} + {\hat{F}}^{m}_{BC} )\, \chi^{m} + 
 \frac{\sqrt{2}}{48}{\rm Tr}\, {\bar{\chi}}^{m} \Gamma^{ABC} \chi^{m} 
 \, \hat{G}_{ABC11} ) \cr 
&&\cr 
 &+&  O(\kappa^{4/3}) \; {\rm relative}\; { \rm to}\; L_S   
\eeqa 
where in eqs(\ref{eq:1},\ref{eq:2}),  
$I = 1, ..11 $ label coordinates on $ M_{11} $; 
 $ A= 1..10 $ those on $M_4 \times X $, ( $a, \bar{b} = 1 ..3$ 
will denote the  
holomorphic and antiholomorphic coordinates on $X$.)  
  The field strength $G_{IJKL}=[ 
 \partial_I C_{JKL} \pm 23 ]\;{\rm terms} + O(\kaps) $ satisfy 
 the modified Bianchi identities \cite{whor} 
 \beq\label{eq:5} 
 (dG)_{11ABCD} = -\frac{3 \sqrt{2}{\kaps}}{\lambda^2} 
 \sum_{m=1}^2 \delta^{(m)} (x^{11})\, [ {\rm Tr}F^{m}_{[AB}\,F^{m}_{CD]} 
 - \frac{1}{2} {\rm Tr} R_{[AB} \, R_{CD]} ] 
 \eeq 
 and in particular one can solve (\ref{eq:5}) by defining  a modified 
 field strength \cite{whor} 
\beq\label{eq:6} 
G_{11ABC} = ( \partial_{11} C_{ABC} \pm 23\;  {\rm terms } + \frac{\kaps}{\sqrt{2} 
\lambda^2} \sum_{m=1}^2 \delta^{(m)} (x^{11}) ( \omega^{(m)}_{ABC} - 
\frac{1}{2} \omega^{(L)}_{ABC} ) 
\eeq 
where $\omega^{(m)} $, and  $ \omega^{(L)} $ are 
($E_8 )$ Yang Mills and Lorentz Chern-Simons 3 forms  defined on the respective 
boundaries, and $\lambda^2 = 2 \pi (4 \pi \kaps )^{2/3} $ is 
the $d=10$ gauge coupling constant. $\kappa = m_{11}^{-9/2} $ , with
$m_{11} $ the 11 dimensional Planck mass. 
 The delta functions $\delta^{(m)} (x^{11}) $ 
have support on the two fixed point sets in $M_4\times X \times {S^1}/ {Z_2} $. 
The presence of these various source terms in  eqs(\ref{eq:5},\ref{eq:6}) 
are an important difference with the corresponding Bianchi identities 
relevant to compactification of the perturbative $E_8 \times E_8 $ 
heterotic string, where $H_{ABC} $ plays the role of $G_{11ABC} $. 
 
In the bulk $d=11$ supergravity lagrangian $L_S$, $g = {\rm det} (g_{IJ})$ 
involves the $d=11$ bulk metric. In the boundary lagrangian, 
the same quantity is understood as being the determinant of the 
$d=10 $ metric obtained as the restriction of the bulk metric to 
 either of the two boundaries $M^{10}_m, m = 1,2 $. Similarly the 
 two copies of the $E_8 $ super Yang Mills fields defined on the boundaries 
 are denoted by $F^{mAB} , \chi^{m} $ respectively. $\Omega_{ABC} $ are the 
 usual $d=10$ spin connections, with hatted quantities denoting the 
 supercovariant generalizations, explicit definitions of which 
 can be found in \cite{whor}. 
 
We shall be interested in compactification of the terms in 
$L_S + L_B $ relevant to the derivation of the effective potential 
$V_{eff}$ obtained when gaugino condensation occurs on one of the 
boundaries corresponding to the hidden sector (which we identify 
as the boundary component $ M^{10}_{m=2} $) in the presence of a non vanishing $G_{11ABC} $. 
In compactifying to $d=4$ we shall adopt a similar procedure discussed in 
\cite{dud} where the sequence is $11\rightarrow 5 \rightarrow 4 $ 
in the bulk action and $10 \rightarrow 4 $ on the boundaries. 
The other subtlety that occurs in M-theory compactification to 
$d =4$ arises because the internal  six dimensional 
metric $g_{ij}, i = 4, ..9 $ is a function in general of the orbifold 
coordinate $x^{11} $ as well as $x^i $. Indeed it was shown in \cite{strw}
 that this 
is a necessary condition in order to get unbroken $N=1 $ supersymmetry in 
four dimensions. In what follows we want to consider the dependence of $V_{eff} $ 
on the moduli associated with the overall scales of $g_{ij} $ and the 
$Z_2 $ invariant metric $g_{11,11}$. Thus in the spirit of \cite{dud} 
we assume we have metrics $g^{(0)}_{ij}( x^{11}, x^i) $ 
 and $g^{(0)}_{11,11} $  which have the right shape but the wrong "size", 
 i.e. we write in compactifying from $d =11 \rightarrow 5 \rightarrow 4$ 
\beqa\label{eq:7} 
 g^{(11)}_{\mu' \nu' } &=& e^{-2 \sigma ( x^{\mu} ) } \, g^{( 5)}_{\mu' \nu'},\quad 
 g^{(11)}_{ij}  =\frac{1}{2} \,  e^{\sigma ( x^{\mu}) } \, g^{(0)}_{ij}(x^{11}, x^{i} ), \quad 
 \cr 
&&\cr 
 g^{(5)}_{11,11} &=& e^{ 2 \gamma (x^{\mu} )} \, g^{(0)}_{11,11} , \quad 
 g^{(5)}_{\mu \nu} = e^{-\gamma(x^{\mu})} \, g^{(4)}_{\mu \nu} 
 \eeqa 
 where   $\mu' = 0,..3, 11 $; $ \mu = 0...3 $ and the 
 superscripts on the metrics in () brackets indicate the particular 
 dimension the metric is defined in. We have emphasised in 
 (\ref{eq:7}) that the fields $\gamma $ and $\sigma$ only depend 
 on the four dimensional coordinates $x^{\mu} $ but each of the 
 metrics $g^{(0)}_{ij} , g^{(0)}_{11,11} $ and $ g^{(4)}_{\mu \nu} $ 
 can depend on  $x^i, x^{11}, x^{\mu} $ in a way determined 
 by the requirement of unbroken  $N=1, d=4 $ supersymmetry \cite{strw} . 
 Our approach thus follows more closely the original one by the 
 authors of \cite{dine} rather than the dimensional truncation approach to 
 N=1, d=4 supersymmetry as employed  
recently in   \cite{dud} ,\cite{lopez} . 
 
The choice of factors of $\sigma $ and $\gamma $ in (\ref{eq:7}) 
yields canonically normalized Einstein-Hilbert actions in d= 4 in the 
supergravity basis \cite{dud}, (i.e. there is a kinetic energy for the 
field $\gamma $ ). 
 
The (curved space) d= 11 gamma matrices satisfy $\{ \Gamma^{I},\, \Gamma^{J} \} 
= 2 g^{(11) IJ} $, and corresponding to the scalings in (\ref{eq:7}) 
we have the following $\sigma$ and $\gamma$ dependence  of their 
various components after compactification : 
 
\beqa \label{eq:8} 
\Gamma_{(11)}^{\mu'} &=& \Gamma_{(5)}^{\mu'} \,e^{\sigma}, \quad  \quad 
\Gamma_{(11)}^i = \Gamma_{(5)}^i \,e^{-\sigma /2}  \cr 
&&\cr 
\{ \Gamma_{(5)}^{\mu'} ,\, \Gamma_{(5)}^{\nu'} \} &=& 2 g^{(5) \mu' \nu'}\,
,\quad 
\{\Gamma_{(5)}^i ,\,  \Gamma_{(5)}^j \} = 2 g^{(0)ij} \,  ,\quad 
 \{\Gamma_{(5)}^{\mu'}, \,  \Gamma_{(5)}^i \} = 0 
\eeqa 
and furthermore 
 
\beqa\label{eq:9} 
\Gamma_{(5)}^{\mu'=11} &=& \Gamma_{(4)}^{\mu'=11} \, \, e^{-\gamma}\, , \quad 
\Gamma_{(5)}^{\mu} = \Gamma_{(4)}^{\mu} \,  e^{\gamma /2}  \cr 
 &&\cr 
\{\Gamma_{(4)}^{\mu} , \, \Gamma_{(4)}^{\nu} \} &=& 2 g^{(4) \mu \nu} \, , \quad 
\{\Gamma_{(4)}^{11} ,\, \Gamma_{(4)}^{11} \} = 2 g^{(0)11,11}\,  , \quad 
\{\Gamma_{(4)}^{\mu}, \,  \Gamma_{(4)}^{11} \} = 0 
\eeqa 
Horava has shown that the combination of terms involving $G_{11ABC}$ 
and $ \chi $ in $L_S$ and $L_B$ can be written as a bulk perfect square 
action $L_{sq} $: 
 
\beq\label{eq:10} 
L_{sq}  =  - \frac{1}{12 \kaps}  \int_{M^{11}} d^{11} x \, \sqrt{g} \,
( G_{11ABC} - \frac{\sqrt{2}}{16 \pi} {(\frac{\kappa}{4 \pi})}^{2/3} 
\delta^{(2)} ( x^{11} )\, {\rm Tr} \, \bar{\chi} \Gamma_{ABC} \chi \,  )^2 
\eeq 
where it should be emphasised that the perfect square term 
should transform as a scalar with respect to $d=11 $ coordinate transformations 
hence a factor of $g^{11,11} $ is implicit in  (\ref{eq:10}). (Recall 
that the delta functions $\delta^{(m)} (x^{11}) $ are not invariant, 
but rather transform as covariant vectors under coordinate transformations 
of $x^{11} $ ). Integrating (\ref{eq:10}) over $x^{11} $ we have: 
 
\beqa\label{eq:11} 
L_{sq} = & -  &\frac{1}{12 \kaps}  \int_{M^{11}} d^{11} x \, \sqrt{g_{(11)}} 
\, ( G_{11ABC} \, G^{ABC}_{11} g^{11, 11} ) \cr 
&&\cr
  &+& \frac{1}{6 \kaps}  
 \frac{\sqrt{2}}{16 \pi} {(\frac{\kappa}{4 \pi})}^{2/3} 
  \int_{M^{10}_2} d^{10} x \, \sqrt{g_{(10)}} \, {\rm Tr} \,  (G_{11ABC} 
\,\, \bar{\chi} 
  \Gamma^{ABC} \chi ) { ( \sqrt{g_{11,11}}) }^{-1}  \cr
&&\cr 
 &-& \frac{1}{12 \kaps}\, (  
 \frac{\sqrt{2}}{16 \pi} {(\frac{\kappa}{4 \pi})}^{2/3}{)}^2 \,
  \int_{M^{10}_2} d^{10} x \, \sqrt{g_{(10)}}\, {\rm Tr}\, (\bar{\chi} \Gamma_{ABC} 
  \chi ) ( \bar{\chi}\Gamma^{ABC} \chi ) \, \delta^{inv}(0) 
  \eeqa 
where $\delta^{(inv)} (x^{11}) \equiv (\sqrt{g_{11,11}})^{-1} \, \delta (x^{11}) $ 
is a delta function transforming as a scalar under $x^{11} $ coordinate 
transformations. 
 
An important feature of M-theory compactification as shown by Witten \cite{strw}, 
is the fact that the volume associated with the compact six dimensional internal 
spaces at  each of the boundaries $M^{10}_i = M^4 \times X_i $ are 
different. The interpolating metrics $g_{ij}(x^i, x^{11}) $ in 
 (\ref{eq:7}) depend on $x^{11}$ in general, and in \cite{strw} it was shown, 
 by considering an expansion about an $x^{11}$ independent metric, that 
 the volumes $V_m$ of $X_m, m =1,2 $ defined at the two fixed point sets $x^{11} = 0 $ and 
 $x^{11} = \pi \rho $ are related:
 
\beqa\label{eq:12}
 V_2(\sigma, \gamma ) &=& V_1 (\sigma , \gamma ) - \frac{\pi}{4 \sqrt{2}} \, \rho\, \int_{X_2} 
d^6 x\, \sqrt{g} \, \alpha (g, F)  \cr  
&&\cr
 &=& V_1 (\sigma , \gamma ) - \frac{\pi}{4\sqrt{2}}\, 
  e^{\gamma} \, m^{-1}_{11} \int_{X_2} d^6 x \, \sqrt{g^{(0)}} \, 
 \alpha (g^{(0)}, F )  
\eeqa 
where $d^6 x \,  \sqrt{g} \alpha (g, F)  =  \frac{\sqrt{2}}{\pi}
 {(\frac{\kappa}{4 \pi} )}^{2/3} \,
\omega \wedge [{\rm Tr} F^{(2)} \wedge F^{(2)} - \frac{1}{2} 
{\rm Tr} R \wedge R ]$ is a 6-form defined wrt the metric 
$g_{ij} $,  ($\omega $ being the K\"ahler form ), 
and $\rho $ is the length of the line element $S^1 / {Z_2} $ with 
respect to the metric choice $g_{11,11} = 1 $ \cite{strw}. 
In obtaining the second line in  (\ref{eq:12}), we have used the 
$\sigma$ and $\gamma $ dependence given in (\ref{eq:7}) and the 
relation $\rho = e^{\gamma - \sigma} m_{11}^{-1} $, which expresses 
the radius  in supergravity units.   
 
The volumes $V_1 (\sigma ,\gamma ) $ and $ V_2 (\sigma, \gamma ) $
define
the four dimensional  observable and hidden sector gauge coupling
constants $g_w $ and $g_h , \footnote{ With the standard embedding of the
spin connection into the gauge connection occuring at the boundary
$X_1 $ the group $E_8 $ is broken to $E_6 $ and the corresponding coupling
$g_w $ is smaller than $g_h $ associated with the boundary $X_2 $ . } $

\beq\label{eq:13}
\frac{1}{{g_w}^2} =  \frac{8 \pi V_1}{({4 \pi \kaps})^{2/3}}
\qquad ,  \qquad 
\frac{1}{{g_h}^2} =  \frac{8 \pi V_2}{({4 \pi \kaps})^{2/3}} 
\eeq 

The result found by Witten  (\ref{eq:12}) translates directly into
the statement that the difference  $g_h^{-2} - g_w^{-2} $ depends only on the
particular integral defined in  (\ref{eq:12}).
Another way of understanding this result, from a weak coupling
perspective was presented by Banks and Dine \cite{banks}.
They pointed out that if the effective four dimensional gauge
couplings are small and one considers a large radius Calabi-Yau space,
the individual moduli dependence of $g_w $ and $g_h $ can be computed.
The result follows directly as a consequence of supersymmetry and holomorphicity
once the axionic couplings are known. Such couplings can be determined from reducing the
$d=10 $ Green-Schwarz terms. For a single overall $(1,1) $ modulus $T$ and
modulus $S$ of the weakly coupled 10 dimensional theory,  the couplings
are given by
\beqa\label{eq:coup}
g_w^{-2} &=& (32 \pi)^{1/3} ( {\rm Re}\, S ) + \frac{m_{11}^2}
{4 (4 \pi)^{1/3} }
 ( {\rm Re }\, T ) \,   \int_{X_2}
d^6 x \, \omega \wedge [{\rm Tr} F^{(1)} \wedge F^{(1)} - \frac{1}{2} 
{\rm Tr} R \wedge R ] \cr
&&\cr
g_h^{-2} &=& (32 \pi)^{1/3}  ( {\rm Re} \, S ) -
\frac{ m_{11}^2 }{4 (4 \pi)^{1/3} }
 ( {\rm Re } \, T ) \, \int_{X_2}
d^6 x \, \omega \wedge [{\rm Tr} F^{(2)} \wedge F^{(2)} - \frac{1}{2} 
{\rm Tr} R \wedge R ] 
\eeqa 
where we emphasise that here $S$ is the weak coupling modulus, whose
real part is related to the dilaton in the usual way.

The question arises whether (\ref{eq:coup}) might be  true even if
one is in a region where the M-theory description is more appropriate.
The answer appears to be affirmative, because  it has been noted in
\cite{whor} that the Chern-Simons term $CGG $ in the action (\ref{eq:1})
turns into an effective Green-Schwarz term, when one uses the relation
\beq\label{eq:GABCD}
G_{ABCD} = - \frac{3 \kappa^2 }{\sqrt{2} \lambda^2 }\, \epsilon (x^{11} )
\, ( {\rm Tr} F_{[AB} F_{CD]} +...... )
\eeq
which solves the modified Bianchi identity
(\ref{eq:5}) near $x^{11} = 0 $. Taking into account both boundaries,
one could once again derive axionic couplings following the arguments of
\cite{banks} and hence  from supersymmetry and holomorphicity,
 conclude that the tree level moduli dependence of
$g_w $ and $g_h $  has the same functional form as in (\ref{eq:coup})
but with $S$ and $T$  moduli given by

\beq\label{eq:21} 
 S = e^{3 \sigma} + i \theta_1 , \quad T = e^{\gamma} + i \theta_2  
\eeq 
 $\theta_1 $ and $\theta_2 $ being related  to the
 axionic degrees of freedom.\footnote{These general conclusions have also been found
in a recent analysis of the four dimensional effective action obtained from
the Horava-Witten 
Lagrangian \cite{low}, where the authors include additional effects due to 
low energy gauge fields.} 

The the idea that certain features of the perturbative gauge coupling threshold corrections can be successfully extrapolated to strong coupling has been discussed
 in \cite{ns}, based on anomaly cancellation arguments.
The form of the couplings in (\ref{eq:coup}) is clearly a manifestaion of this and 
has been used recently in the discussion of the 
pattern of the soft supersymmetry breaking 
masses in M-theory 
\cite{noy}.

In studying the mechanism of gaugino condensation,
we shall assume that the boundary $X_2 $ has non contractible 
cycles which allow one to turn on  Wilson lines corresponding to the 
the hidden sector 
$E_8 $, breaking the latter to some group $G$ whose coupling constant becomes 
strong at some scale $\mu $. We want to determine the dependence of 
$\mu $ on $\gamma$ and $\sigma $. General arguments  relate the scale 
$\mu $ to the GUT scale $M_{GUT} $ which is in turn given by the 
masses of the gauge mesons corresponding to the  broken generators in the 
process $E_8 \rightarrow G $.

Considering the compactification of the pure Yang-Mills action on 
$M^{10}_2 $ to $d=4 $ the relevant terms are  
 
\beq\label{eq:14} 
S^{(4)}_{YM   } = - \frac{1}{4}( \frac{1}{4 \pi {g_h}^2} )\, \int d^4 x \, ( {\rm Tr} 
\, F_{\mu \nu }F^{\mu \nu} 
+ e^{- 3\sigma - \gamma}\, {\rm Tr}\, D_{\mu}A_i D^{\mu} A^i  
+ ....)  
\eeq 
Wilson lines correspond to the scalars $A_i $ transforming in the adjoint 
of $E_8$ acquiring non vanishing expectation values. The scale of 
$\langle A^i \rangle  $ can be determined by the requirement that metric 
independent integral   $\int_{\Gamma} \langle A \rangle \cdot d{\bf x} $ 
(where $\Gamma $ is a non-contractible cycle in $X_2 $) is of order 
unity. This then implies that $\langle A \rangle $ is typically of order 
${m^{-1}}_{11} $. To determine the gauge meson masses, we have to 
rescale the   $d=4$ fields $A_\mu  $ and $A_i $ appearing in 
(\ref{eq:14}) to obtain canonical kinetic energies. Thus we define 
the canonical fields $\tilde{A}_{\mu} ,\, \tilde{A}_i $ 
\beq\label{eq:95} 
A_{\mu} = 2 \sqrt{\pi} \, g_h \,  {\tilde{A}}_{\mu},  \qquad 
A_i = 2 \sqrt{\pi} \, g_h \,  e^{3 \sigma /2 + \gamma /2}\, {\tilde{A}}_i 
\eeq 
from which it follows that,  
\beq\label{eq:15} 
M_{GUT} \approx M_{meson} \approx  m_{11} \, e^{-3 \sigma /2 - \gamma
/2}, \quad   
\quad  \mu (\sigma, \gamma) \approx M_{GUT} \,  e^{- 1/ 2 b_0{g_h}^2 (\sigma ,\gamma )  } 
\eeq 
where $b_0$ is the coefficient of the first term in the 
beta function associated with the coupling $g_h $, and 
$\mu $ is defined as the scale at which  the running coupling 
$g_h $ becomes strong. In the perturbative case \cite{dine}, there is a similar 
expression for $\mu $ except there the multiplicative factor is 
$e^{-2 \sigma} $ only, and  the M-theory mass scale $m_{11} $ 
is replaced by the $d=4$ Planck mass $M_{pl} $. 
 
In order to discuss hidden sector gaugino condensation, we need also to 
define canonical kinetic terms for the $d=4$ gauginos associated 
with the gauge group $G$ of the strongly coupled sector. 
Here by canonical we mean that the  gauginos are normalized in the same 
way as the corresponding gauge fields, as required by unbroken 
$N=1$ supersymmetry.  Using the metric ansatz of (\ref{eq:7}), 
and the various rescalings of the gamma matrices given in 
(\ref{eq:8},\ref{eq:9}) the kinetic energy for the  massless 
$d =4$ gauginos of the group $G$ are  
 
\beq\label{eq:16} 
 \int d^4 x \, \{ - \frac{1}{2 {g_h}^2 } \, e^{-2 \sigma -\gamma } 
 {\rm Tr} \, \bar{\chi} \gamma^{\mu} D_{\mu} \chi  \, \} 
 \eeq 
from which we learn that the correctly normalized four dimensional gaugino 
$\chi_{(4)} $ = $\sqrt{2}\,  e^{-\sigma -\gamma /2} \,  \chi $. 
With this normalization we expect that the strongly interacting gauge 
group $G$ induces nonvanishing expectation values for gaugino 
bilinears at a mass scale determined by $\mu (\sigma, \gamma ) $ i.e. 
$ |\langle {\bar{\chi}}_{(4)} \chi_{(4)}  \rangle | \sim \mu^3 (\sigma, \gamma ) $. 
 
In principle, to obtain the supersymmetric Wilsonian action below
the scale $\mu$ one should integrate correctly 
all massive modes from $m_{11} $ down to $\mu$ in a supersymmetry preserving way. 
Assuming this can be done, one ends up 
with a supersymmetric effective action.  Here we adopt a practical approach and 
employ the simplified procedure of Dine et al. \cite{dine}. 

Now we want to derive an effective potential from the terms in 
(\ref{eq:11}) whose origin was the perfect square term in (\ref{eq:10}). 
The procedure we shall adopt is to replace gaugino bilinears in terms of the 
quantity $\mu ^3 (\sigma, \gamma ) $ allowing at the same time complex phases 
in the latter. Whether the perfect square structure, which is apparent in 
$d=11$ survives compactification would seem to depend on the 
precise form we take for $G_{11ijk} $ and the bilinears 
$< \bar{\chi}\Gamma_{ijk} \chi > $ . The situation appears more complicated 
than in the perturbative case. There for example instead of $G_{11ijk} $ 
we have the 3 form $H = d B  + \omega^{YM} - \omega^{L} $, 
the latter terms being the difference between Yang- Mills  ($E_8 \times E_8 $ ) 
and Lorentz Chern-Simons 3-forms. Although the possibility of turning on these latter terms was 
raised in \cite{dine}, the usual procedure is to allow dirac like string singularities 
in the potential $B_{ij} $ such that $H_{ijk} = c \epsilon_{ijk} $, where 
$c$ is a constant. (Here we momentarily change our notaion so that $i$
refers to holomorphic internal indices).
 Then the 3-form $H $ is closed. This is consistent 
with the Bianchi identities if the topological condition 
$\int {\rm Tr} F \wedge F - {\rm Tr} R \wedge R = 0 $ holds. 
This constraint can be achieved by the usual embedding of the spin connection 
in the gauge connection, which  also implies the vanishing of the 
Chern-Simons terms in  $H_{ijk} $. At the same time general arguments lead 
one to take $< \bar\chi \Gamma_{ijk} \chi > $ to also be proportional to 
the components $ \epsilon_{ijk} $ of the holomorphic 3 form on the 
internal Calabi-Yau manifold. (This follows because massless d=4 
gauginos in this case, are associated with spinors on the Calabi-Yau space  
that are singlets under the $SU(3) $ holonomy group). The result of this is that the 
perfect square structure of the potential persists in $d=4 $. 
 
In the case of M-theory, $G_{11ABC} $ as given in (\ref{eq:6}), 
is analogous to $H_{ABC} $ in the perturbative case, except that the 
 Chern-Simons term has been ``split apart'',  
and has support only at the $Z_2 $ fixed points 
 sets. The importance of this as pointed out in \cite{strw} , is that 
 while the standard embedding is sufficient to allow a solution to the 
 ($ d=11 $) Bianchi identities, (since this involves integrating the 
 right hand side of (\ref{eq:5}), which picks up both fixed point 
 contributions ), it is not sufficient for pointwise cancellation of the 
 Yang-Mills and Lorentz Chern-Simons terms in $G_{11ABC}$. As has been 
 suggested \cite{strw}, this might imply that  $G_{11ABC} \neq 0 $ generically 
 in M-theory compactifications. 
 
 Thus it would appear that the Chern-Simons source terms in $G_{11ABC} $ 
 are something one should take into consideration when calculating 
 the effective action. Additionally, one could of course consider turning on 
 the $C_{11ij} , C_{ijk} $ potentials, and to this extent the 
 perfect square form of the potential written as a  $d=11$ integral 
 (\ref{eq:6}) suggests that the latter potentials are such that 
 $G_{11ijk}  $ develops a vev localized on the strongly coupled 
 boundary, to compensate the similarly localized gaugino condensate \cite{horava}. 
 We can in principle take into account both these possibilities by 
 taking $G_{11ijk} = \sum_{m=1}^2 \omega^{(m)}_{ijk}(x^i) \delta^{(m)} (x^{11} ) $, 
 where $\omega^{(m)}_{ijk} $ could represent either the Chern-Simons terms 
 or taking $\omega^{(1)}_{abc} = 0,\, \omega^{(2)}_{abc} =
 c \, m^3_{11}  \epsilon_{abc} $
 we would get an expression similar to the perturbative case.

 Similarly, the issue arises as to the form of the fermion bilinear 
 $\langle \bar{\chi} \Gamma_{ijk} \chi \rangle $. For now we shall be equally general 
 and define  $ {\rm Tr} \,\langle \bar{\chi} \Gamma_{ijk} \chi  \rangle  =
  \mu^3 \, {\tilde{\omega}}_{ijk}( x^i ) $  describing
   a condensate localized only on the boundary 
   $X_2 $. 
 
   With these ansatzes the naive effective potential obtained from 
   (\ref{eq:11}) can be computed. Before we do this, the obvious 
   difficulties associated with the presence of singular terms 
   involving $\delta (0) $ have to be considered. These issues were 
   raised in \cite{whor} where their contributions to the $d=11$ supergravity 
   action with boundaries was first obtained. In general it is 
   believed that these 
   are artifacts of the singular nature of the boundary and its purely 
   classical description. Quantum M-theory effects may serve to regularize 
   such singularities by thickening the boundaries, a typical length scale 
   being the  M-theory scale $m_{11}^{-1} $. In what follows 
   we shall adopt this approach,  and replace the 
   invariant delta function  $\delta^{inv} (0) $  with 
   $a \, m_{11}^{-1} $, where the arbitrary constant $a$  serves simply to 
   rescale the effective action in $d=4$. Hence for simplicity  we set $a =1$.
 
   Taking into account the various normalizations and scalings 
   discussed earlier, we arrive at the following form for the 
   effective potential: 
 
   \beqa\label{eq:17} 
   V_{eff}(\sigma , \gamma ) &  =  & \frac{m_{11}^4 }{96 \pi^2 (4 \pi{)}^{4/3}} 
   \, e^{-3 \sigma - 3\gamma } \, \{ \, B_1 + B_2 + 2 B_3 \, \mu^3 (\sigma, \gamma ) 
   \, e^{ 9\sigma /2 + 3 \gamma /2 } \cr 
   && \cr 
   & + & B_4  \, \mu^6 (\sigma , \gamma ) \, e^{9 \sigma + 3 \gamma} \,  \} 
   \eeqa 
 
   where the quantities $B_1 .... B_4 $ are given by the following 
   integrals 
 
   \beqa\label{eq:18} 
     B_1 &=& \int_{X_1} d^6 x \, \sqrt{g^{(0)} }\, \omega^{(1)}_{ijk} 
     \,\, \omega^{(1)}_{i'j'k'}  g^{(0) ii'}  g^{(0) jj'}   g^{(0) kk'} \cr  
    &&\cr 
     B_2 &= &\int_{X_2} d^6 x  \, \sqrt{g^{(0)} } \,\omega^{(2)}_{ijk}\,\, 
   \omega^{(2)}_{i'j'k'}  g^{(0) ii'}  g^{(0) jj'}   g^{(0) kk'} \cr 
   &&\cr 
    B_3 &=& \int_{X_2} d^6 x \,  \sqrt{g^{(0)} } \, \omega^{(2)}_{ijk} \,\,
   {\tilde{\omega}}_{i'j'k'}  g^{(0) ii'}  g^{(0) jj'}   g^{(0) kk'} \cr 
   &&\cr 
    B_4 &=& \int_{X_2} d^6 x  \, \sqrt{g^{(0)} } \, {\tilde{\omega}}_{ijk}\,\, 
   {\tilde{\omega}}_{i'j'k'}  g^{(0) ii'}  g^{(0) jj'}   g^{(0) kk'}  
    \eeqa 
 
   In the above integrals, the scaling factors of 
   $e^{\sigma } $ have been taken out, and it is understood that the metrics 
   $g^{(0)}_{ij}$ are those obtained from the bulk metrics 
   restricted to the appropriate boundary. 
 
   Similar integrals arise in the perturbative case, the difference being 
   that there is a single  Calabi-Yau space $K$ and a single 
   metric (so for example  $B_1 $ is absent and $K$ can be identified 
   with $X_2$. In that case the 3 forms entering the integrals are all 
    proportional to the holomorphic 3-form on $K$ \cite{dine} , 
    and hence the integrals $B_2 .. B_4 $ are all proportional to 
    each other with coefficients that reproduce a perfect square 
    structure in $ V_{eff} $.    
 
   In the present case, there are further subtleties. This is 
   because $B_2 .. B_4 $ involve integrals over the boundary component 
   $X_2  $ located at the fixed point set $x^{11} = \pi \rho $ 
   whilst $B_1 $ is defined with respect to $X_1 $. In principle there 
   could be a hidden dependence in $B_2 ... B_4 $ on the 
   moduli $\sigma , \gamma $ analogous to the moduli dependence 
   of the volume integral at $X_2 $. One can try and determine this dependence 
   using similar methods as in \cite{strw}. In doing this one can define the 
   integrals $B_2( x^{11} ), ...B_4(x^{11} ) $ at an arbitrary value of 
   $x^{11} $, and obtain a differential equations for them by 
   taking $g^{(0)}_{a \bar{b} }(x^{11}, x^i) =  g^{(0)}_{a \bar{b}}(x^i) + 
   h^{(0)}_{a\bar{b} }(x^i , x^{11} ) + ... $. (It should be noted that 
   here the perturbation in the metric $  $ is $e^{-\sigma}  $ times the 
   quantity $ h_{a\bar{b} }(x^i , x^{11} ) $ defined in \cite{strw} .) 
 
   In deforming the integrals away from the boundary $X_2 $ the 
   only dependence on $x^{11} $ occurs through the various metric 
   factors, since the ansatz for $\omega^{(m)} $, and $\tilde{\omega} $ 
   implies these are $x^{11} $ independent. 
 
   Using the expressions for $\partial_{11} h_{a\bar{b} } $ 
   derived in \cite{strw} 
   one can obtain the following equations satisfied by $B_2 ... B_4 $ 
   away from $X_2 $ 
 
   \beqa\label{eq:19} 
   \partial_{11} B_2 (x^{11}) & =& \frac{3}{4 \sqrt{2}}\, e^{\sigma}\, 
    \int_{X} d^6 x \, 
    \sqrt{g^{(0)} } \,\omega^{(2)}_{ijk} \, 
   \omega^{(2)}_{i'j'k'}  g^{(0) ii'}  g^{(0) jj'}   g^{(0) kk'} 
   \alpha (g^{(0)} , F)  \cr 
    &&\cr 
   \partial_{11} B_3 (x^{11}) & =&\frac{3}{4 \sqrt{2}} \, e^{\sigma} \, \int_{X} 
    d^6 x\, 
    \sqrt{g^{(0)} }\, \omega^{(2)}_{ijk} \, 
   {\tilde{\omega}}_{i'j'k'}  g^{(0) ii'}  g^{(0) jj'}   g^{(0) kk'} 
   \alpha (g^{(0)} , F)  \cr 
    &&\cr 
   \partial_{11} B_4 (x^{11}) & =&\frac{3}{4 \sqrt{2}} \, e^{\sigma} \, \int_{X} 
    d^6 x \, 
    \sqrt{g^{(0)} }\, {\tilde{\omega}}_{ijk} \, 
   {\tilde{\omega}}_{i'j'k'}  g^{(0) ii'}  g^{(0) jj'}   g^{(0) kk'} 
   \alpha (g^{(0)} , F)  
  \eeqa 
 
  Next we can derive expressions for $\partial^{2}_{11} B_{\alpha} (x^{11} ) ,\, 
  \alpha = 1..4 $. This will involve $\partial_{11} \alpha $  
  an expression for which can be determined through the Bianchi identities 
  of $G_{11ABC} $ without the source terms  \cite{whor}, with the result that 
  it is a total derivative wrt $x^i $.  In fact 
  $\partial_{11} \alpha \sim \omega^{ij} \omega^{kl} (dG)_{11ijkl} $, 
  where $(dG)_{11ijkl} $ only involves derivatives wrt the coordinates 
  $x^i $, and again it should be stressed that here $G_{11ijk} $ 
  contains no source terms, unlike that defined in  (\ref{eq:5}). 
  In the context of gaugino condensation,  if such terms are turned on 
  at all they are given in terms of the holomorphic (and antiholomorphic) 
  3-forms, which are covariantly constant and hence  in this 
  specific case, $\alpha $ is $x^{11} $ independent. 
   Consequently, 
   $\partial^{2}_{11} B_2 (x^{11} ) =  \partial^{2}_{11} B_3 
   (x^{11} )= \partial^{2}_{11} B_4 (x^{11} ) = 0 $, i.e. the integrals 
    defined in 
    (\ref{eq:19}) are $x^{11} $ independent to this order. 
    This situation is now similar to the volume calculation of $X$, 
    and one has the following expressions for the original 
    integrals (\ref{eq:18}) 
 
    \beqa\label{eq:20} 
    B_2 & = & B_1 - \frac{3}{4 \sqrt{2}} \pi \rho \, e^{\sigma} \, ( 
    \int_{X_2} d^6 x \, \sqrt{g^{(0)} } \, {( \omega^{(2)} )}^2 
      \alpha (g^{(0)} ) ) \cr 
    &&\cr 
    B_3 & = & B^{(1)}_3  - \frac{3}{4 \sqrt{2}} \pi \rho \, e^{\sigma} \, ( 
    \int_{X_2} d^6 x \, \sqrt{g^{(0)} } \,{ \omega^{(2)} \cdot \tilde{\omega} } \,
      \alpha (g^{(0)} ) ) \cr 
   &&\cr 
    B_4 & = &  B^{(1)}_4 - \frac{3}{4 \sqrt{2}} \pi \rho \, e^{\sigma} \, ( 
    \int_{X_2} d^6 x \, \sqrt{g^{(0)} } \, {( \tilde{\omega}) }^2 
      \alpha (g^{(0)} ) ) 
   \eeqa 
where in (\ref{eq:20}), the integrals $B^{(1)}_3 $ and $B^{(1)}_4 $ are  
defined at the fixed point set boundary $X_1 $, and so together with  
$B_1 $ are independent of the moduli $\sigma , \gamma $. Indeed this also 
applies to the integrals over $X_2 $ in (\ref{eq:20}), the only dependence  
on the moduli being through the combination $ \rho \, e^{\sigma} $, which  
from our previous definitions is proportional to $e^{\gamma} $. 
 
At this point we can see that the integrals $B_2 ....B_4 $ have a  
dependence  on $\gamma $ that is not present in the perturbative case. 
(The coefficients in perturbative case would correspond to just keeping  
integrals $B_2 , B_3 $ and $B_4 $ and keeping just the first terms  
on the right hand side of (\ref{eq:20} ), and identifying $X_1 $ with  
the space $K$ in \cite{dine}  endowed with metric $g^{(0)}_{ij} $,  
and setting $\omega^{(2)}_{a bc},\, \tilde{\omega} $ proportional to  
the holomorphic 3 form $\epsilon_{abc} $ on $K$. This reproduces the perfect square 
potential.)  
 
It is hard to see how this perfect square structure holds if, as we  
discussed earlier, we really take into account the source terms  
which would appear to be present in $G_{11abc} $ 
at $x^{11} = 0 $ and $x^{11} = \pi \rho $, because then if ${\tilde{\omega}}_{abc} $  
was taken to be proportional to $\epsilon_{abc} $, $B_2, B_3 $ and $B_4 $ 
are not obviously related to each other. If however we somehow ignore these  
source terms,  and $G_{11abc} $ develops a piece purely at the boundary $X_2 $ as  
advocated in \cite{horava}, and again proportional to $\epsilon_{abc} $, then  
the square structure survives. However, even if this is so the presence  
of moduli dependent terms in $B_2 ... B_4 $ may not be consistent with  
the existence of a superpotential. We shall discuss these issues in the 
next section. 

\section{Effective superpotential in M-theory gaugino condensation}
In this section we shall, following the reasoning used in the perturbative
approach \cite{dine}, attempt to identify an $N=1, d=4$ superpotential $W(\sigma,\gamma ) $ which can reproduce, in some approximation, the above scalar 
potential. Before we do this we have to identify the $d=4$ K\"ahler structure 
that emerges in our compactification i.e. define complex moduli fields, 
whose  real parts will be related to $\sigma , \gamma $ and  their 
corresponding K\"ahler potential. This has been studied by various authors   
\cite{dud} , \cite{lopez} in the dimensional truncation approach, but is equally applicable in our case (just as the identification of moduli, and K\"ahler potential in  
the perturbative case \cite{dine} was carried out in more general context than 
dimensional truncation). 

The K\"ahler potential  of the $S$ and $T$ moduli defined  earlier
in (\ref{eq:21}) is \cite{dud}
 
\beq\label{eq:22} 
K = - {\rm ln} ( S + \bar{S} ) - 3 {\rm ln} ( T + \bar{T} )  
\eeq 
Now in  order to extract a superpotential we have to be careful in  
considering what the effective $d=4 $ Planck mass is after compactification 
as this enters the well known formula for the scalar potential in $N=1, 
d=4$ supergravity.

We have followed the conventions in \cite{dud} regarding the various $\sigma $ and $\gamma $ 
dependent scalings in the metric. After compactification 
one obtains   an $N=1,
d=4$ supergravity in the  `supergravity basis' where the $\gamma $ field
has a non-vanishing kinetic energy, and the $d=4$ Einstein action is 
canonically normalized (i.e. all factors of $e^{\gamma} $ and $e^{\sigma }$
cancel in front of the $d = 4 $ curvature term.).
In these units the effective four dimensional Planck mass is of order
$m_{11} $, and is the basis for which the usual text book 
formulae for the effective potential is written
(see \cite{banks} for a discussion of this point.) This is to be contrasted with the theory as expressed in
M-theory units, in which $\gamma  $ has no kinetic term and
there is a reduced Planck mass  $m_{pl}^2 = t \, m_{11}^2 $  \cite{dud}.

With these points in mind, we take the following form for the superpotential
$W(S, T) $ 

\beqa\label{eq:23} 
W(S, T ) & = & m_{11}^3 \{ \,  c_1 + c_2 \,
 e^ {-\frac{3 { ( 4 \pi )}^{1/3}}{b_0}
 {\displaystyle ( S +  c_3 T ) } } \, \} \cr
&&\cr
&= &  m^3_{11} \, c_1 + \tilde{W} (S,T)
\eeqa
where $c_1, c_2 $ are complex coefficients which we want to determine  by
trying to match the potential coming from (\ref{eq:23}) with the one
obtained from compactification in (\ref{eq:17}). The form of the exponent
in $ (\ref{eq:23} ) $ is obtained by expressing the hidden sector coupling
constant  $g_h $ in terms of $S$ and $T$
as given in (\ref{eq:coup}), and from which the constant $c_3$ can
easily be deduced.
  
Superpotentials of the form (\ref{eq:23}) have already been advocated as
being relevant to gaugino condensation in M-theory \cite{banks} 
their general forms following from requirement of holomorphicity and 
shift invariance. The new twist in the expression for $W(S,T) $ , compared
to the perturbative case, is the presence of the $T$ modulus. Clearly this
will give rise to new terms in the scalar potential compared to that case.

Even before we discuss the connection between the scalar potential 
arising from (\ref{eq:23} ) and $V_{eff}$, there are other potential
difficulties associated with the superpotential. 
Recall the way the expectation value of gaugino 
bilinears  $\langle \lambda^a  \lambda^b \rangle $ enters 
the scalar potential of the 4d supergravity \cite{nfer}.
(In  this discussion we adopt the notation where
 $\lambda^a $ denotes the components of the gaugino, $i$ labels complex
moduli fields, and $a$ is an adjoint group index.) 
With canonically normalized (in 4d) gravitational part and gauge and gaugino 
kinetic terms the relevant part of the Lagrangian is
\beq
V  = \; e^K g^{i \bar{j}} 
(D_i W + \frac{1}{4} e^{-K/2} \partial_i f_{ab} \langle \lambda^a 
\lambda^b \rangle ) (D_{\bar{j}} W + \frac{1}{4} e^{-K/2} \partial_{\bar{j}} f_{ab}
 \langle \bar{\lambda}^a 
\bar{\lambda}^b  \rangle )+ \; ...
\label{gbil}
\eeq
where $g^{i \bar{j}}$ is the inverse K\"ahler metric and rest of the
notation is standard (see \cite{nfer} ). Now, let us assume there is no 
perturbative superpotential and switching on 
condensates is equivalent to switching on the nonperturbative superpotential.
Then one should require 
\beq
D_i W_{np} = \frac{1}{4} e^{-K} \partial_i f_{ab} \langle \lambda^a 
\lambda^b \rangle
\label{wcon}
\eeq
However, in the present case the tree-level gauge kinetic function $f_{ab}$ depends on both superfields $S$ and $T$. Hence, 
without specifying the exact form of $W_{np}$ one can obtain 
from (\ref{wcon}) the relation    (assuming that $ \langle \lambda^a 
\lambda^b \rangle $ is non vanishing on the boundary $X_2 $ only, 
with $f$ defined to be the $E_8$ gauge kinetic function) 
\beq
\frac{ \partial_S \log W_{np} - \frac{1}{S+\bar{S}}}{\partial_T \log W_{np} - 
\frac{3}{T+\bar{T}}}= \frac{\partial_S f}{\partial_T f}
\label{constr}
\eeq
The relation (\ref{constr}) is incompatible with assumed holomorphicity 
of $W$ and $f$. The way to understand this relation in the context of 
Horava-Witten model is to recall that the dependence of the gauge coupling on $T$ 
comes in from the terms which, as pointed out in \cite{strw}, can be seen 
as the perturbation of the Lagrangian defined on the boundary with the weakly 
coupled $E_8$. This reasoning would imply that we are not allowed to write 
the troublesome relation (\ref{constr}) as it compares terms which arise in 
different order of perturbative expansion. However, the puzzle is still around, as the same corrections tell us what is the difference between gauge couplings at both 
boundaries, and obviously enter the expression for the hidden $E_8$ condensation
scale, and, as a consequence, any expression for $W_{np}$ we can arrive at.
Of course, this particular problem with holomorphicity 
doesn't imply on its own that 
there is something wrong with supersymmetry - even 
standard 4d supergravity with loop 
corrections in general cannot be formulated in the canonical form known from
standard tree level formulation. In what follows we shall encounter 
further trouble with holomorphicity, arising specifically in M-theory model 
we have here.  

Returning to  the calculation of the scalar potential, 
 $W(S,T) $  together with the K\"ahler potential $K(S,T) $ given in (\ref{eq:22}) gives the following 

\beqa\label{eq:24}
V_{eff} &= & m^4_{11} \, e ^{- 3 \sigma - 3 \gamma } \, \{ e^{6 \sigma}  \vert \,
 ( e^{- 3 \sigma} \,
\frac{W(S,T)}{m^3_{11} } + 6 \,c_2 \, \frac{{(4 \pi )}^{1/3}}{b_0} \,
 e^{ -\frac{3 { ( 4 \pi )}^{1/3}}{b_0} {\displaystyle ( S + c_3 T )} } )\, {\vert}^2 \cr
&&\cr
&+& 3  \, e^{2 \gamma } \,  \vert \, ( e^{-  \gamma} \,
\frac{W(S,T)}{m^3_{11} } + 6 \, c_2 \, c_3 \, \frac{{(4 \pi )}^{1/3}}{b_0} \,
 e^{ -\frac{3 {(4 \pi )}^{1/3}}{b_0} {\displaystyle ( S + c_3 T )} } ) \, {\vert}^2  \cr
&&\cr
&-& 3 \, \vert \,
\frac{W(S,T)}{m^3_{11}} \,{\vert}^2 \}
\eeqa
  
Using the expression for $g_h $ in terms of $S$ and $T$, one can rewrite
(\ref{eq:24}) in  a more suggestive manner

\beqa\label{eq:25}
V_{eff} & = & m^4_{11} \, e^{-3 ( \sigma + \gamma ) } \, \{ \,  
\vert \, c_1 + ( 1 + \frac{3}{2 b_0 g^2_h }\, ) \, \frac{\tilde{W}(S,T)}
{m^3_{11}}\, {\vert }^2 \cr
&&\cr
&-& [ \, 6 \, c^2_3 \, \frac{{ ( 4 \pi )}^{2/3}}{b^2_0 \, g^2_h} \, (T + \bar{T} ) \,
 + 12 \, c_3 \, \frac{{( 4 \pi )}^{2/3}}{b^2_0}\,
(S + \bar{S} )( T + \bar{T} ) \, ] \, \frac{ \vert \tilde{W} {\vert}^2 }
{m^6_{11} }  \}
\eeqa

At this point it is interesting to compare the form of $V_{eff} $ in 
(\ref{eq:25}) with the one obtained in compactifications of the 
perturbative heterotic string with hidden sector gaugino condensation
and $H_{ijk} $ field strength turned on \cite{dine}. Here, in the simplest scenario,
there are also two moduli $S$ and $T$ related to the dilaton $\phi $ and 
scale factor $e^{\sigma} $ via ${\rm Re}S = e^{3 \sigma } \, {\phi}^{-3/4} $ 
and ${\rm Re} T = e^{\sigma } \,  \phi^{3/4} $.  The relevant
superpotential has the same form as in (\ref{eq:23} ) except that the 
term in $T$ in the exponential is absent. 
One finds an effective potential $V^{(p)}_{eff} $ given by 

\beq\label{eq:26}
V^{(p)}_{eff} = \frac{1}{16} M^4_{pl} \, e^{- 6 \sigma } \, \phi^{-3/2}
\, \vert \, c + h \, [  (\frac{3}{2 b_0} ) ( S + \bar{S} ) + 1 ] \,\, 
e^{- \frac{3S}{2 b_0} } \,\, {\vert }^2 
\eeq 
This potential reproduces the perfect square form obtained explicitly 
from compactification only up to the power law correction term 
proportional to $( S + \bar{S} ) $ inside [  ] in (\ref{eq:26}). 
In this  theory, the four dimensional gauge coupling $g^{-2}  = {\rm Re}
S$, so such a correction is of order  $3/( b_0 \, g^2 )  $.

Now we return to the $V_{eff} $ in the M-theory case. It is clear that
there are two corrections terms present, proportional to $1/ g^2_h $, 
but in addition there is a third term (the last term in 
 (\ref{eq:25} ) ) which is not a correction in this
sense, and which represents a deviation away from a perfect
 square form of the potential. This is a direct consequence of having both
an $S$ and $T$ dependence in the superpotential. 
 Thus we should in principal find such terms 
in our naive compactification  effective potential in (\ref{eq:17}), 
along with  perfect square terms.

In fact this lack of perfect square structure fits in with our earlier 
observations concerning $V_{eff} $ in (\ref{eq:17}). To try and match 
(\ref{eq:25}) with (\ref{eq:17}), let us consider the more likely
possibility discussed earlier, namely
 $G_{abc}  = {(4 \pi )}^{5/3}  c\, m^3_{11} \, \epsilon_{abc} \,\delta ( x^{11} - \pi \rho ) $, 
and $ {\rm Tr} \, \langle \,{\bar{\chi}}_{(4)} \Gamma_{abc} \chi_{(4)}\, \rangle =
 {(4 \pi )}^{5/3} h \, \mu^3
\,\epsilon_{abc} $,with $c$ real and $h$ complex. Then only the integrals 
$B_2 ... B_4 $ are relevant in (\ref{eq:17}). The moduli independent
parts of these integrals  then reproduce the perfect square term in 
(\ref{eq:25}) with
 $\vert c_1 \vert^2 =  c^2 \,m^6_{11} \,\int_{X_2}d^6 x \, (\, \sqrt{g^{(0)} }\, \epsilon^2\,)
 $ and $\vert c_2 \vert^2 = \vert h {\vert}^2 \, m^6_{11} \,  \int_{X_2}d^6 x \, (\, \sqrt{g^{(0)} }\,
\epsilon^2 \, )  $. One can expect that  $ m^6_{11} \,\int_{X_2}d^6 x \, (\,
\sqrt{g^{(0)} }\,\epsilon^2 \, ) $ will be of order $1$  since it involves the metric
$g^{(0)}_{ij} $.

However, there are additional moduli dependent pieces to $B_2 ...B_4 $,
which are proportional to $e^{\gamma } $. Interestingly, the 
additional factors $(S + \bar{S} )(T + \bar{T} ) $ found (\ref{eq:25})
 have also this $\gamma $ dependence, but there
are difficulties. Firstly the additional terms in (\ref{eq:25}) 
are proportional to $\vert \, \tilde{W} \, {\vert}^2 $ and not 
$\vert \, W \, {\vert}^2 $ as implied by (\ref{eq:17}). Secondly
 the coefficients  do not match
up precisely, (e.g. the factors of $b_0 $ in (\ref{eq:25}) are absent in
the integrals $B_2 ... B_4 $.) Nevertheless the coefficients do resemble 
each other in that one has $\int_{X_2 } d^6 x \, \sqrt{g^{(0)} }  
\alpha (g^{(0)}, F ) $ in $c_3 $ and $\int_{X_2 } d^6 x \, \sqrt{g^{(0)} }  
\, \epsilon^2 \, \alpha (g^{(0)}, F ) $ occuring in the second terms of 
$B_2 ... B_4 $ (\ref{eq:20}).

Clearly these difficulties are related to those raised earlier
in connection with having tree level gage kinetic functions that 
are dependent on both $S $  and $T$. Certainly the  troublesome 
moduli dependent factors in $B_2 ... B_4 $ are a consequence of the 
same phenomenon that lead to a tree level dependence on $T$ 
in the hidden sector gauge coupling.  
                                                   
One should add that the situation is somewhat worse if we entertain 
the possibility of  turning on 
the Chern-Simons forms  in $G_{11abc} $ at each of the two 
$Z_2 $ fixed points. In such an event, it does not seem likely that
$B_1 ... B_4 $ have the structure necessary to reproduce any perfect
square terms. It is hard to be  definitive because in this case it 
is difficult to obtain an explicit dependence of these integrals on \
the coordinate $x^{11} $. This is because the terms we are turning on in 
$G_{11abc} $ are not covariantly constant so the analysis that lead to 
(\ref{eq:20}) is not obviously applicable.

Putting all these various problems aside, we can at least try to make some
preliminary
observations concerning the possible stabilization of the moduli 
expectation values.
One of the motivations in trying to derive an effective potential from
having both $G_{11abc} $ and $\langle  \bar{\chi} \chi \rangle $ non vanishing was 
to hopefully avoid the usual runaway problem concerning the values 
of $S$ and $T$ at the minimum of $V_{eff} $.  Banks and Dine 
\cite{banks}  have 
already discussed the situation in the context of M-theory, 
at least for 
superpotentials of the form (\ref{eq:23}) but with $c_1 = 0 $, i.e.
they do not consider the more general situation of 
allowing non-vanishing $G_{11abc} $. Translating this to our potential
(\ref{eq:17} ), it appears possible that one might achieve stabilization of 
$\sigma $ and $\gamma $ (and hence the radius $\rho $ ) if one 
turns on e.g. $G_{11abc} $ at the boundary $X_2 $ only and 
proportional to $\epsilon_{abc} $, which was the first scenario 
discussed earlier. Although we do not present details here, what appears to
be  important in possibly achieving this
stabilization is the relative minus signs of the moduli dependent
parts of $B_2...B_4 $ compared to the moduli independent parts.

Finally, before ending this section, we comment on the results we have 
presented here, and the effective superpotentials obtained by 
\cite{dud} when applying the Scherk-Schwarz compactification procedure in 
the context of M-theory defined on $M_4 \times X \times  S^1 /Z_2 $.
Such a comparison is motivated by the ideas presented  in \cite{anton},  that 
the Scherk-Schwarz mechanism applied in this way is effectively 
equivalent to gaugino condensation. One of the key features of the induced
superpotentials in \cite{dud} was that the axionic shift symmetry associated 
with the modulus $S$ is violated.   
We have found no evidence of this from  the form of the effective potential we
have obtained in (\ref{eq:17} , \ref{eq:20} ). Furthermore if one added
a linear term in $S$ to the superpotential (\ref{eq:23}), (to simulate the 
term found in \cite{dud} ), there are  additional terms, some of which   
 will violate $S$ shift symmetry  in the corresponding 
scalar potential that again we have no evidence for in  
(\ref{eq:17} ,\ref{eq:20}).

\section{Soft terms} 
In this section we make some  preliminary comments on the pattern of soft  
masses in the observable sector.
 At tree level in the observable sector the no-scale structure 
appears which is well known already from the weakly coupled heterotic string  
models. Since the observable superpotential is of no-scale type (trilinear  
terms in matter fields) the dependence of the scalar potential on  
matter fields $C$  relevant for the calculation of masses is of the form  
\beq 
V=V(T + \bar{T} - |C|^2)  
\eeq 
which implies  
\beq 
\frac{\partial^2 V}{\partial_C \partial_{\bar{C}}}|_{min, C=0}= 
-\frac{\partial V}{\partial T}|_{min, C=0} = 0 
\eeq 
However, there is no exact symmetry of the full Lagrangian which could  
prevent scalar masses from arising through radiative corrections  
(supersymmetry is broken and massive gravitino couples to observable fields). 
Similarly, gaugino masses will be generated through radiative corrections.  
In this respect the situation is similar to that in no-scale models 
studied in the past \cite{ellis}. However, one can speculate  
that in some sense gaugino masses can arise even at tree level. 
The inverse gauge coupling can be regarded as the real part of the  
 $x^{11}$-dependent holomorphic function $f$  of $S$ and $T$ superfields 
\beq 
f= \alpha_1 S  + \beta_1 T - T \Delta (x^{11}) 
\eeq 
where  $\alpha_1 $ and $\beta_1 $ are constants 
( $\beta_1 $ is immediately given by (\ref{eq:coup}) )and  $\Delta (x^{11}=0)=0$. In reality one expects that in quantum M-theory  
the singularities associated with the zero-thickness of the boundary are  
regularized by thickening the boundaries,  
the effective thickness being of the  
order of $m_{11}^{-1}$, as discussed earlier.
 In this case the resulting $\Delta_{eff} (0)
\approx \Delta (m_{11}^{-1}) $ can be different from zero. Now, in 4d supergravity the tree-level gaugino masses  
are proportional to $f_s F^S + f_T F^T$. One doesn't know at this point the  
relation of the $F^{S,T}$-terms at $x^{11}=0$ to those which we can compute  
at the strongly coupled boundary using the effective Lagrangian derived here, 
but these quantities computed at the strongly coupled boundary in general 
 don't have to both  vanish.  
 Of course, one can say that there exist M-theory models 
 which are similar to weakly coupled string models, in these models 
one would expect in analogy T-dependent threshold corrections and future 
interesting structure in ``twisted'' sectors, but in this paper we 
want to restrict ourselves to the specific model at hand and to separate 
facts from speculations.

\section{Conclusions}
In this paper we have derived the  explicit form, and discussed properties, of 
 moduli dependent effective potentials
as arising from compactification of M-theory when one of the boundaries 
supports a strongly interacting gauge sector and induces 
gaugino condensation. 
Consideration has been given to allowing non vanishing components of 
the three form field strength $G_{11ABC} $. We have found that 
the naive process of trying to match the potential of moduli obtained 
by compactifying terms in the  M- theory action, with the potential
obtained from  the superpotential $W(S,T) $ given in (\ref{eq:25})
is problematic. The origin of these difficulties appears to be due to 
the 'tree level' dependence of the hidden sector gauge coupling on the 
moduli $S$ and $T$.  As we have argued in the section 3, although one 
may try and make sense of this by  arguing that the $S$ and $T$ dependence
might be from different orders in perturbation theory, this is at odds
with  what one obtains from compactification. Such difficulties are 
not apparent when the same ideas are employed in the perturbative 
heterotic string. It could be that these problems are an indication that 
in considering gaugino condensation in M-theory, and  not carefully
integrating
 out massive modes in obtaining an effective  theory at scales
  $\ll m_{11} $, (rather using a more  naive approach which worked in the
 perturbative case),leads to difficulties. Perhaps a related point is to    
understand 
deeper issues concerning the effective four dimensional theory obtained
from M-theory for example the connection between Wilsonian and physical 
gauge couplings, which has revealed  many subtle issues in the past \cite{kl}
(and references therein ).  
\vskip 0.25cm

\subsection*{Acknowledgments} 
The work of Z.L. was partially supported
by Polish Commitee for Scientific Research grant 2 P03B 040 12, and 
by M.Curie-Sklodowska Foundation and Polish-French Collaborative Projects 
Programme. The work of S.T. was supported by the Royal Society of Great Britain.


\begin{thebibliography}{000} 

\bibitem{polch} J. Polchinski, {\em String Duality}, eprint  hep-th/9607050.
\bibitem{strw} E. Witten, Nucl. Phys. B 471 (1996) 135.
\bibitem{whor} P. Horava, E. Witten, Nucl. Phys. B 475 (1996) 94;
 P. Horava, E. Witten, Nucl. Phys. B 475 (1996) 94.
\bibitem{revs}  J.P. Derendinger,
L.E. Ib\'a\~nez, H.P. Nilles, Phys.Lett. {\bf 155B} (1985) 467;
; T.R. Taylor,
Phys.Lett. {\bf 164B} (1985) 43.
\bibitem{dine}  M. Dine, R. Rohm, N. Seiberg, E. Witten, Phys.Lett. {\bf 156B} (1985) 55.
\bibitem{horava} P. Horava, Phys. Rev. D 54 (1996) 7561.
\bibitem{anton} I. Antoniadis, M. Quiros, {\it Phys. Lett.} {\bf B392}
(1997) 61; hep-th/9705037 ({\it Nucl. Phys. B} to appear; 
 preprint CERN-TH/97-165, CPTH-S548.0797, IEM-FT-160/97 , 
eprint hep-th/9707208.
\bibitem{dud} E. Dudas, C. Grojean, preprint CERN-TH/97-79,
LPTHE-ORSAY 97/17 Saclay T97/038, eprint hep-th/9704177.
\bibitem{lopez} T. Li, J.L. Lopez and D.V. Nanopoulos, 
hep-ph/9702237; hep-ph/9704247. 
\bibitem{banks} T. Banks and M. Dine {\it Nucl.Phys.} {\bf B479} (1996), 173.
\bibitem{nfer}   S. Ferrara, L.
Girardello, H.P. Nilles, Phys.Lett. {\bf 125B} (1983) 457.
\bibitem{ellis} J. Ellis, D.V. Nanopoulos, M. Quiros and 
F. Zwirner, {\it Phys. Lett.} {\bf B180} (1986) 83;
 J. Ellis, A.B. Lahanas, D.V. Nanopoulos, M. Quiros and 
F. Zwirner, {\it Phys. Lett.} {\bf B188} (1987) 408.
\bibitem{kl}  V. Kaplunovsky and J. Louis, 
{\it Nucl.Phys.} {\bf B422} (1994), 57; 
{\it Nucl.Phys.} {\bf B444} (1995), 191.
\bibitem{ns} H.P. Nilles and S. Stieberger, 
  {\it Nucl.Phys.}  {\bf B499 } (1997), 3.
\bibitem{noy} H.P. Nilles, M. Olechowski and M. Yamaguchi, 
preprint TUM-HEP-282/97, SFB-375/201, eprint hep-th/9707143 . 
\bibitem{low} A. Lukas, B.A. Ovrut and D. Waldram,
preprint UPR-771T, PUPT-1723, HUB-EP-97/51, hep-th/9710208. 
 
\end{thebibliography}
\end{document}